\documentclass[twocolumn,amsmath,amssymb, showpacs, showkeys]{revtex4}

\usepackage{graphicx}
\usepackage{dcolumn}
\usepackage{bm}

\begin{document}


\preprint{APS/123-QED}

\title{Spectroscopy of a Cooper-Pair box in the Autler-Townes configuration}

\author{E. J. Griffith}
 \email{e.j.griffith@liverpool.ac.uk}
\author{J. F. Ralph}
 \affiliation{
 Department of Electrical Engineering and Electronics,\\ The University of Liverpool, Brownlow Hill, Liverpool, L69 3GJ, United Kingdom.}

\author{Andrew D. Greentree}
\affiliation{
The Centre for Quantum Computer Technology, School of Physics,\\ The University of Melbourne, Victoria 3010, Australia.}

\author{T. D. Clark}
\affiliation{
Centre for Physical Electronics and Quantum Technology,\\ The University of Sussex, Falmer, Brighton, BN1 9QT, United Kingdom.}

\date{\today}

\begin{abstract}
\textbf{Abstract:} A theoretical spectroscopic analysis of a microwave driven superconducting charge qubit (Cooper pair box) coupled to an LC oscillator model is performed.  By treating the oscillator as a probe through the backreaction effect of the qubit on the oscillator circuit, we extract frequency splitting features analogous to the Autler-Townes effect from quantum optics, thereby extending the analogies between superconducting and quantum optical phenomenology.  These features are found in a frequency band that avoids the need for high frequency measurement systems and therefore may be of use in qubit characterisation and novel coupling schemes.  In addition we find this frequency band can be adjusted to suit an experimental frequency regime by changing the oscillator frequency.\end{abstract}



\maketitle


\section{\label{sec:sec1}INTRODUCTION}

Superconducting charge qubits (Cooper pair boxes) are promising candidates for use in quantum information processing and quantum computing systems.  Many of the necessary results for quantum computing have already been demonstrated in these systems including single qubit rotations \cite{nakamura:nature, paladino:qubit}, single-shot readout \cite{Astafiev:PRB} and two-qubit entanglement \cite{pashkin:nature}.  

In common with other solid-state technologies (e.g. Si:P \cite{Kane:Nature}), superconducting devices are expected to be highly scalable due to the ability to construct large numbers of qubits using existing lithographic manufacturing~\cite{you:scalable}.  However, any lithographic procedure has intrinsic manufacturing tolerances and there can be significant variation in device parameters between neighbouring qubits.  These are of little consequence in classical digital logic designs because the `zero' and `one' logical states are usually well separated in terms of the levels used to represent them \cite{gibson:book}.  They are well-defined discrimination thresholds and they are highly robust to small fluctuations.   Such classical error protection fails as the number of particles (electrons) defining a given state is reduced to the few particle level and the fragility of quantum coherence introduces new problems.  Hence the Hamiltonian of each individual qubit will need to be characterized to unprecedented levels \cite{Schirmer:Hamiltonian, Cole:Hamiltonian}.

Recently, attention has turned to the coupling of superconducting flux and charge qubits to microwave oscillators and resonators  \cite{delft:oscillator, Johansson:Rabi, Buisson:entangled, Averin:MQCBook, Wei:bus}.   These will have some resonant frequencies caused by the intrinsic capacitance and inductances in the connection path, this can often be approximated by linear oscillators for small fluctuations and deviations.  The simplest example of this is in the usual spectroscopy of the qubit where the microwave field acts as a weak probe \cite{Nakamura:PRL1997}, and increasing the strength of the field allows Rabi oscillations \cite{Nakamura:Rabi}.  To investigate the strong-coupling limit of cavity Quantum Electrodynamics, Wallraff \textit{et al.} constructed the analog of an optical cavity for a Cooper-Pair Box, and showed single photon dynamics \cite{Wallraff:Nature}.  This work and related theoretical analysis \cite{Brandes:OpticsReport} clearly suggests that the full canon of quantum optics \cite{Mandel:OpticsBook} can be translated to the superconducting arena for potential technological advantage.  In addition, the ability to fabricate desired pseudo-atoms allows investigation of phenomena (particularly multi-state phenomena) that would not be convenient using real atoms.

We have investigated a simulated superconducting charge qubit model, controlled via a classical microwave pump and a bias/control voltage.  We treat the bias control as a fully quantized linear oscillator with a finite resistance (RLC oscillator) \cite{ferry:book} which is coupled to a thermal environment \cite{spiller:qsd}.  The quantised oscillator is used to allow the field to represent non-adiabatic processes and entanglement between the qubit and bias field.  



	\begin{figure*}
		\centering
		\includegraphics[width=0.75\textwidth]{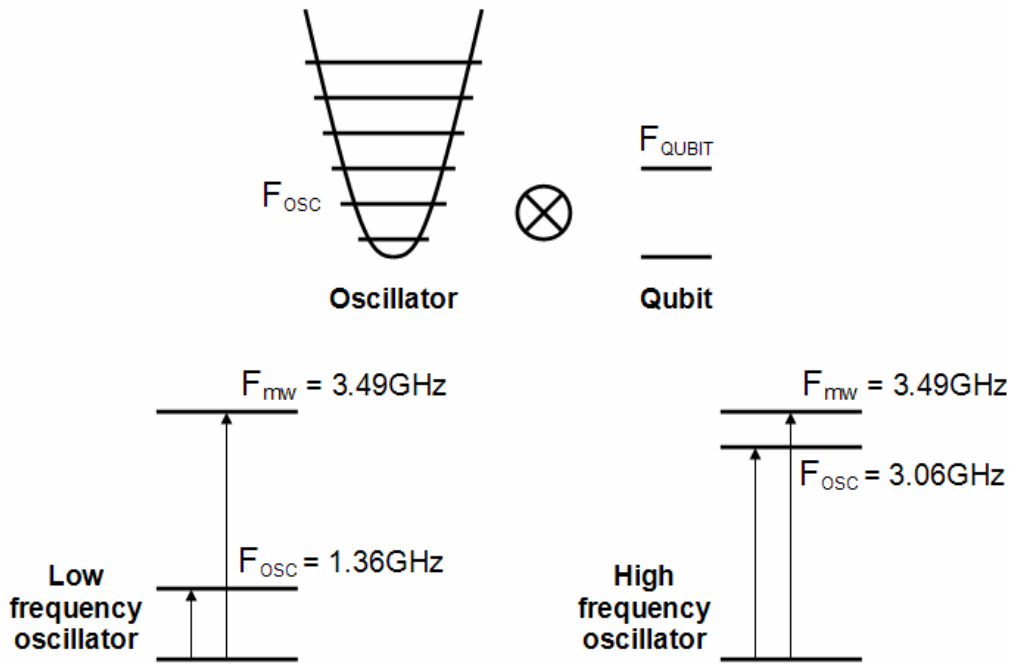}
		\caption{\label{fig:QubitOscEnergy} A two level qubit is coupled to a many level harmonic oscillator, investigated for two different oscillator energies.  Firstly, the oscillator resonant frequency is set to 1.36GHz, this more resembles the conventional configuration such that the fundamental component of the oscillator does not drive the qubit.  However, we also investigate the use of a high frequency oscillator of 3.06GHz which can excite this qubit.  In addition, qubit is constantly driven by a microwave field at 3.49GHz to generate Rabi oscillations and in this paper we examine the relation between these three fields.}
		\label{fig:QubitOscEnergy}
	\end{figure*}	

System measurements can be performed on the oscillator circuit or the qubit, but measurement of the oscillator is generally preferred as it is more readily accessible via weak measurements than the fragile qubit state.  One of the ways to reduce decoherence is to reduce the number of degrees of freedom coupling directly to the qubit, thereby limiting the environmental noise.  In addition, the oscillator is of particular interest as its evolution is significantly affected by the qubit dynamics due to backaction caused by the two way capacitive qubit-oscillator coupling.

This paper examines the power spectral density (PSD or noise spectrum) of the oscillator and qubit as functions of the two main control parameters, microwave amplitude and bias voltage.  The PSD, particularly for low sub GHz frequencies, is a readily accessible experimental measurement \cite{Sillanpaa:reactive, Ilichev:monitor}.  By exploring non-standard parameters for the bias circuitry and microwave control, our investigations show a rich spectroscopy of dynamically controllable features. Beyond fundamental investigations, these results may be of benefit in the characterisation of qubit parameters, and the tunability of the features may allow for new control techniques for single and multiple qubit coupling~\cite{ralph:guidance}.  These features are not restricted to superconductive devices, as semiconductor charge qubits may also exhibit similar behaviour.  The magnitude of the features are of a smiliar order to the oscillator peak and so should be detectable if the oscillator peak can be sucessfully measured. 

We consider two situations (Fig.~\ref{fig:QubitOscEnergy}) in which we change the resonant frequency of the biasing circuit, we have chosen the values to be non-commensurate as to avoid the possiblity of higher oscillator harmonics repeating the same effects. Initially,  we examine a conventional configuration in which the bias oscillator  frequency ($f_{osc} = 1.36$GHz) is much less than both the undriven qubit transition frequency ($f_{qubit} = 1.94$GHz) and the microwave drive frequency ($f_{mw} = 3.49$GHz).  This oscillator frequency is too slow to create excitations, so only a single set of microwave driven excitations (Rabi oscillations) are observed.
The second scenario, involves using a high frequency oscillator ($f_{osc} = 3.06$GHz) \cite{irish:osc} near the qubit and microwave frequencies.  This smaller detuning creates a range of interesting features in the oscillator spectrum.  
 We also observe a frequency splitting when the qubit is correctly biased in the presence of a driving signal and strong coupling.  This resembles the Autler-Townes effect observed in quantum optics \cite{AutlerTownes:PR}.  In this case, the oscillator frequency is sufficient to create its own excitations, which intersects and mixes with the microwave driven Rabi oscillations.  This mixing allows the Rabi frequency to be modulated/upconverted into an experimentally favourable frequency band, in addition we find that the position of this band is can be adjusted by changing the oscillator frequency.

The Autler-Townes (A-T) experiment is a well-known experimental scheme from quantum optics for probing a strongly driven transition.  In the canonical example \cite{AutlerTownes:PR}, an atomic transition is driven by a strong, quasi-resonant field.  This induces Rabi oscillations, and due to decoherence processes, and if the driving is resonant, the population in the two states will tend to equalise.  In this case the most usual way to describe the system is in the dressed-state basis, where the dressed states are the eigenstates of the driven system, and correspond to orthogonal coherent superpositions of ground and excited state. The transient evolution of a three-state system in an A-T-style configuration is studied experimentally and theoretically in \cite{Greentree:threestate}.

Because the population in ground and excited state tends to equalise in the long time limit, the fidelity of a readout signal can be compromised.  Against this backdrop, the A-T configuration provides an additional readout handle.  By sweeping a weak probe field, quasi-resonant with a transition from the one state to a third, otherwise uncoupled state, and ensuring that the sweep range covers all of the substructure of the dressed states induced by the strong coupling field, the population in the dressed states, rather than the bare atomic states is read out.  As this is a more natural basis to be monitoring in a strongly driven system anyway, the experiment allows a useful measurement of otherwise hidden dynamics to be performed.  Good examples of systems well resolved by this process are bichromatic \cite{greentree:autler} and polychromatically driven transitions \cite{Greentree:PRA99, Papademetriou:Bichro}, and V-systems \cite{Echaniz:PRA}.  In general because the third state is assumed unaffected by the strong driving, then the population difference between the dressed states and the third state can, in principle, be very large, and is unaffected by the strong coupling. 
  
More generally, however, the A-T experiment provides for more flexibility, and introduces an alternative measurement handle.  In this context it has potential utility in solid-state systems, where the Rabi oscillations are replaced with coherent-tunnelling oscillations, and the tuning of transitions is performed electrically via surface gates.  In this paradigm, schemes related to the A-T experiment have been proposed as alternative readout mechanisms for a charge qubit operating in the superposition basis \cite{Greentree:PRB} and for singlet-triplet discrimination in a two-spin system \cite{Greentree:twospin}.  As will be shown below, the increased flexibility of the A-T configuration in our system emphasises side-bands corresponding to subharmonic resonances \cite{Agarwal:PRA1986} and features at multiples of the Rabi frequency frequency \cite{Windsor:PRA}, which would otherwise be hidden in more conventional, two-state systems, and afford new diagnostic methods, and possibilities for control.



 	\begin{figure*}
		\centering
			\includegraphics[width=0.81\textwidth]{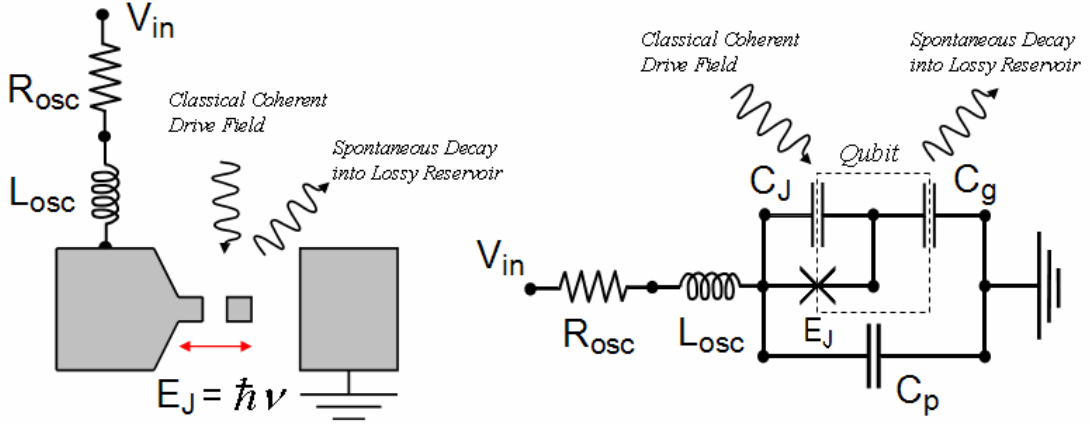}  
	  	\caption{\label{fig:SystemModel} An abstract layout of a simple single junction Cooper pair box, with the biasing voltage connection to the first filter stage modelled by a resistor and inductor.   The Cooper pair box is modeled by an equivalent circuit network, consisting of three capacitors: $C_J$ - the capacitance of the Josephson junction, $C_g$ - The capacitance of the island gate, and $C_p$ - the parasitic capacitance formed between the biasing electrodes.  These capacitors, together with the resistor and inductor form an RLC oscillator circuit, the inductance value is changed to investigate the two  oscillator frequencies, (1.36GHz and 3.06GHz).}
			\label{fig:SystemModel}
	\end{figure*}

\section{\label{sec:sec2}SYSTEM MODEL}

The system model (shown in Fig.~\ref{fig:SystemModel}) is a small superconducting island, capacitively coupled to a bulk superconductor via a Josephson junction, allowing Cooper pairs to coherently tunnel on and off the island.  The Josephson junction is sufficiently weak and the total capacitance of the island sufficiently small that the energy associated with charging the island is large compared to the tunnelling energy of the junction ($\hbar\nu$), this is normally called the Coulomb blockade regime \cite{aleiner:blockade}.  The necessary biasing voltage is applied through the bulk superconductor, across the Josephson junction.  In addition, the qubit is also coupled capacitively to a grounding electrode, a reference point for the junction biasing voltage.

The bias circuit has natural intrinsic dynamics with a characteristic frequency response, we approximate this by a resonant RLC oscillator in series with the island, which takes into account the intrinsic inductance and capacitance present in the bias circuit, as well as a parasitic capacitance across the bias electrodes.  This oscillator is modelled quantum mechanically as a standard $N$-state Simple Harmonic Oscillator (SHO) allowing non-adiabatic processes and not limiting the analysis to low frequency fields \cite{ralph:flux,griffith:charge,ralph:charge,spiller:classosc}.  The bias oscillator is also connected to an external amplifier through which the oscillator voltage can eventually be measured.  We have focused on the effects of components immediately coupling to the qubit, although it should be acknowledged that the specific entire chain \cite{bladh:chain} which will add more resistive, capacitive and inductive loads.  The qubit can be indirectly measured using its reactive backreaction on the oscillator voltage, the backreaction is a result of the two-way capacitive coupling between the qubit and oscillator circuit.  Indeed, the presence of fluctuations in the oscillator will also have an effect on the qubit, an additional source of environmental decoherence, this is minimised by maintaining a suitably low temperature (10mK).

The Hamiltonian for this coupled system is:	
	\begin{eqnarray}
		\label{eq:H}
		H & = & \frac{q^2}{2C_q} +  \kappa\frac{qQ}{C_{q}} + \frac{Q^2}{2C_Q} + V_{in}Q \nonumber \\
		  & + & \frac{\Phi^2}{2L_{osc}} - \hbar\nu\sin\left(2\pi\frac{\theta}{\Phi_0}\right) 
	\end{eqnarray}	

\noindent with $Q$ and $q$ being the effective oscillator and island charges respectively, $\Phi$ is the inductor flux and $\theta$ the superconducting phase difference.  the effective island and bias electrode capacitances and coupling parameter $\kappa$ are given by:	

\begin{subequations}
	\label{eq:Caps}
  \begin{eqnarray}
  	\label{eq:CapsA}
		C_q & = & \frac{C_gC_J + C_JC_p + C_pC_g}{C_g + C_p}  
		\\
		\label{eq:CapsB}
		C_{qQ} & = & \frac{C_gC_J + C_JC_p + C_pC_g}{2C_g}   
		\\
		\label{eq:CapsC}
		C_Q & = & \frac{C_gC_J + C_JC_p + C_pC_g}{C_g + C_J}  
		\\
		\label{eq:CapsD}
		\kappa & = & \frac{C_q}{2C_{qQ}}	
  \end{eqnarray}
\end{subequations}

\begin{table} [h]
	\caption{\label{tab:components}Components (Capacitors same for both $f_{osc}$)}
	\begin{ruledtabular}
		\begin{tabular}{clc}    
			 & Description & Value\\
			\hline
			$\nu/2\pi$ 	& Josephson energy 										& 1.94GHz\\
			$C_J$ 			& Josephson junction capacitance 			& 0.5fF\\
			$C_g$ 			& Qubit-Grounded Bulk capacitance			& 0.5fF\\
			$C_p$ 			& Electrodes parasitic capacitance 		& 10pF\\
			\hline
			$L_{osc}$ 	& Bias inductance ($f_{osc}=1.36$GHz)	& 1.37nH\\
			$R_{osc}$ 	& Bias resistance ($f_{osc}=1.36$GHz)	& 60m$\Omega$\\
			$\gamma$		& Qubit jump rate ($f_{osc}=1.36$GHz)	& $6.8\times10^7$\\
		  $T1$				& Relaxation time ($f_{osc}=1.36$GHz)	& 14.7ns\\
			\hline
			$L_{osc}$ 	& Bias inductance ($f_{osc}=3.06$GHz)	& 0.27nH\\
			$R_{osc}$ 	& Bias resistance ($f_{osc}=3.06$GHz)	& 27m$\Omega$\\
			$\gamma$		& Qubit jump rate ($f_{osc}=3.06$GHz)	& $1.53\times10^8$\\
		  $T1$				& Relaxation time ($f_{osc}=3.06$GHz)	& 6.5ns\\
			\hline
			$\kappa$ 		& Coupling parameter 									& $5\times10^{-5}$\\
			$Q_D$ 			& Oscillator quality factor 					& 200\\
		\end{tabular}
	\end{ruledtabular}
\end{table}



The qubit geometry is transformed to a classical electrical circuit model (Fig.~\ref{fig:SystemModel}).
The circuit includes the intrinsic qubit capacitances $C_J$ and $C_g$, (the Josephson
junction capacitance and a grounded bulk connection capacitance).  In addition, a large parasitic capacitance $C_p$ is formed between the bulk superconductor electrodes and is connected to the inductive and resistive elements of the biasing circuit.

The Hamiltonian is derived from consideration of circuit voltage and current laws using a detailed loop analysis method proposed by Burkard \cite{burkard:circuit}, with the Josephson junction current given by the well known phase-current relation (\ref{eq:JJcurrent}) where $\theta$ is the superconducting phase difference between the adjoining segments.
\begin{eqnarray}
	\label{eq:JJcurrent}
	I_J = I_C\sin\left(2\pi\frac{\theta}{\Phi_0}\right)  
\end{eqnarray}

The resulting Hamiltonian (\ref{eq:H}) is rewritten to separate the qubit ($\hat{q}$) and oscillator ($\hat{Q}$) degrees of freedom, and a coupling constant $\kappa$ is introduced.  The $\kappa$ term is typically much less than one, and in the limit $\kappa \longrightarrow 0$ the cross-coupling $qQ$ term is zero and the systems evolve independently. 

In addition, it should be noted that the capacitances $C_J, C_g, C_p$ have been redefined as effective qubit and bias oscillator capacitances, $C_q, C_{qQ}, C_Q$, (\ref{eq:Caps}).  The term $V_{in}Q$ is the contribution of the bias/control voltage and the inductive term arises from the magnetic flux associated with the oscillator geometry. 

Starting from Eq. (\ref{eq:H}) it is possible to group the Hamiltonian terms into three
parts: the oscillator and qubit Hamiltonians, plus a coupling term.
\begin{eqnarray}
	\label{eq:Htensor}
	H = H_{osc}\otimes I_{qubit} + I_{osc}\otimes H_{qubit} + H_{coupling}
\end{eqnarray}

\noindent 
where I is the corresponding identity matrix for each component. 
The Hamiltonian is written in the matrix representation, using
a basis formed from the tensor product of the uncoupled oscillator energy eigenstates and the charge states of the isolated qubit (which
forms the compuational basis for the qubit). The three terms in the Hamiltonian are presented below in Eqs. (\ref{eq:HoscMN},~\ref{eq:HqubitMN},~\ref{eq:HcouplingMN}).  In the these equations, the subscripts are used to label the basis states 
of the uncoupled systems: the $N$-state oscillator is indicated by a subscript `1' and the qubit by a subscript `2'.  
So that $m_1$ or $n_1$ represent the oscillator states 
($m_1, n_1\in \left\{0, 1, 2, ..., N\right\}$) and $m_2$ or $n_2$ indicate the qubit states ($m_2, n_2\in \left\{0, 1\right\}$).
\\

\begin{enumerate}

\item Oscillator Hamiltonian, $H_{osc}$
	
	The oscillator is quantised as a simple harmonic oscillator (\ref{eq:Hosc}) \cite{Buisson:entangled}, with energy $\omega$.
	In the electrical circuit model,	the charge $\hat{Q}$ and flux $\hat{\Phi}$ form the conjugate variables, analogous to momentum $(\hat{p})$ and position $(\hat{x})$, (\ref{eq:HoscA}).  The usual mechanical model of the harmonic oscillator can be equated with the electrical Hamiltonian, yielding equivalent values for an effective `mass' $m = C_Q$ and `spring constant' $k = 1/L_{osc}$.  We also assume if the qubit is maintained at a suitably low temperature (10mK) the contributions of the higher energy levels are negligible, hence we constrain the system to the lowest $N=10$ oscillator states for computational efficiency, whilst checking that probabilities are not lost to the higher states.
	\begin{subequations}
		\label{eq:Hosc}
		\begin{eqnarray}
		  \label{eq:HoscA}
			H_{osc} =  \frac{\hat{\Phi}^2}{2L_{osc}} + \frac{\hat{Q}^2}{2C_Q}~\equiv~\frac{1}{2}k\hat{x}^2 + \frac{\hat{p}^2}{2m}
			\\
			\label{eq:HoscB}
			E_n  =  \hbar\omega\left(n + \frac{1}{2}\right), ~~~\omega  = \sqrt{\frac{1}{L_{osc}C_Q}}
		\end{eqnarray}
	\end{subequations}
	
\noindent tensoring with the qubit basis yields Eq. (\ref{eq:HoscMN}), the first part of the combined system matrix (\ref{eq:Htensor})
	\begin{eqnarray}
	  \label{eq:HoscMN}
		& \left\langle m_1, m_2 \right|& H_{osc}\left| n_1, n_2 \right\rangle = \nonumber\\
		& ~ &\hbar\omega\left(n_1 + \frac{1}{2}\right)\delta_{m_1,n_1}\delta_{m_2,n_2}
	\end{eqnarray}
	where, in this paper, $\omega/2\pi$ is 1.36GHz for the low frequency oscillator with 3.06GHz being the high oscillator frequency.
\\
	
	\item Qubit Hamiltonian, $H_{qubit}$
	
	The qubit basis is defined as the presence or absence of a single Cooper pair on the island, $0e$ and $2e$.  Although there may be higher states present, we assume the probability of a third state is negligible
	\begin{subequations}
		\label{eq:Hqubit}
		\begin{eqnarray}
		  \label{eq:HqubitA}
			H_{qubit} & = & \frac{\hat{q}^2}{2C_q} - \hbar\nu\cos\left(2\pi\frac{\hat{\theta}}{\Phi_0}\right)
			\\
			\label{eq:HqubitB}
			H_{qubit} & = & \frac{2e^2}{C_q} \left|1\right\rangle\left\langle1\right| - \frac{\hbar\nu}{2}\left(            \left|0\right\rangle\left\langle1\right| + \left|1\right\rangle\left\langle0\right| \right)		
		\end{eqnarray}
	\end{subequations}
	
\noindent where the cosine term can be written as the sum of two exponentials divided by two, these exponentials are raising and lowering operators on charge states, which are then expressed in the matrix formulation as off diagonal terms.  Tensoring this small matrix with the oscillator basis yields Eq. (\ref{eq:HqubitMN}), the second part of the combined system matrix (\ref{eq:Htensor}). 
	
	\begin{eqnarray}
		\label{eq:HqubitMN}
		 	& & ~~~~\left\langle m_1, m_2 \right| H_{qubit}\left| n_1, n_2 \right\rangle = \\
		 	& &\delta_{m_1,n_1}  \left(\frac{2e^2}{C_q} \delta_{m_2,1} \delta_{n_2,1} -\frac{\hbar\nu}{2}  \left( \delta_{m_2,1} \delta_{n_2,0} + \delta_{m_2,0} \delta_{n_2,1} \right)\right) \nonumber
	\end{eqnarray}
	
	The biasing field caused by voltage $V_{in}$ can be represented as an effective bias charge $Q_{bias}$.  The microwave field is applied as an effective qubit charge, the combination is $q_{ext}$ \cite{gunnarsson:ext}.
	\begin{subequations}
		\label{eq:qext}
		\begin{eqnarray}
		  \label{eq:qextA}
			q_{ext}\left(t\right) & = & Q_{bias} + A_{mw}\sin\left(\omega_{mw}t\right)
			\\
			\label{eq:qextB}
			\hat{q}\left(t\right) & = & q_{ext}\left|0\right\rangle \left\langle0\right|+\left(q_{ext}+2e\right)\left|1\right\rangle \left\langle1\right| 
		\end{eqnarray}
	\end{subequations}
\\
	
	\item Coupling term, $H_{coupling}$
	
	The qubit-oscillator coupling term is the product of the qubit and oscillator charge operators.
	To quantise this we express the oscillator charge $\hat{Q}$ in terms of raising and lowering operators~\cite{ferry:book}.
	The terms $\sqrt{n_1}$ and $\sqrt{n_1+1}$ are normalisations for the changes in oscillator state. 		
	\begin{subequations}
		\label{eq:Hcoupling}
		\begin{eqnarray}
		  \label{eq:HcouplingA}
			H_{coupling} & = & \kappa \frac{\hat{q}\hat{Q}}{C_q}
			\\
			\label{eq:HcouplingB}
			\hat{Q} & = & -i\sqrt{\frac{\hbar m \omega}{2}} \left(a - a^\dagger\right)	
		\end{eqnarray}
	\end{subequations}	
	
\noindent the tensor product of the $\hat{q}$ and $\hat{Q}$ matrices yields Eq. (\ref{eq:HcouplingMN}), the final part of the combined system matrix (\ref{eq:Htensor}).
	
	\begin{eqnarray}
	  \label{eq:HcouplingMN}
	  & \left\langle m_1, m_2 \right|& H_{qubit}\left| n_1, n_2 \right\rangle = \nonumber\\
	  & & -i\frac{\kappa e}{C_q}\sqrt{2\hbar m \omega}~\delta_{m_2,1}\delta_{n_2,1} \nonumber\\
    & & \left(\sqrt{n_1}~\delta_{m_1,n-1}-\sqrt{n_1+1}~\delta_{m_1,n+1}\right)
	\end{eqnarray}
		
\end{enumerate}



\section{\label{sec:sec3} UNRAVELLINGS AND WEAK MEASUREMENTS}

The evolution of the quantum system (qubit and oscillator) in the presence of environmental fluctuations is described by the usual master equation.

\begin{eqnarray}
   \label{eq:master}
	 \frac{d\rho}{dt} & = & - i \left[H,\rho \right] \nonumber\\
	 & & + \sum_{m} \left( L_m \rho L^{\dagger}_{m} - \frac{1}{2} L^{\dagger}_{m} L_m \rho - \frac{1}{2} \rho L^{\dagger}_{m} L_m \right) 						
\end{eqnarray}
where the ensemble average of the individual state evolutions yield the density operator for the combined system (The qubit and oscillator circuit) $\rho = E\left(\left|\psi\right\rangle\left\langle\psi\right|\right)$.
Eq. (\ref{eq:master}) is the general equation for the average evolution of an ensemble \cite{scully:book}, for which the Caldeira-Leggett model \cite{Caldeira:decohere} is a specific example.  The unravellings represent the evolution of specific systems (not ensemble evolutions) under the actions of projective measurements on their environment.  Here we have ignored the effects of expicit decoherence due to dephasing in individual qubits.

Unravelling the Master equation allows us to represent only the stochastic effects within the qubit and oscillator as a whole.  In this section we describe two unravellings which correspond to spontaneous emissions and the thermal diffusion mechanism of the oscillator.  This environment also includes the effects of an external measuring system on the oscillator as this will add to the decoherence.  We have chosen a quantum state diffusion (QSD) unravelling (\ref{eq:QSDstate}) to model oscillator dissipation, although equally valid is the quantum jump (QJ) unravelling representing a different measurement process on the environment, indeed both unravellings should agree on average \cite{brun:QSDandQJ}.  


\subsection{\label{sec:sec3pA} Oscillator quantum state diffusion}

Quantum state diffusion is applied to simulate the effects of a finite temperature environment, and dissipation \cite{spiller:qsd}.  The diffusion model consists of random thermal and quantum fluctuations combined with a progressive drift of the quantum state.  The system stochastic evolution
is now governed by Eq. (\ref{eq:QSDstate}).
The first term is the usual Schr\"{o}dinger evolution, the second term is deterministic and represents the drift, and the final term models the random fluctuations of the state.  The intensity of these effects is governed by the Lindblad operators defined in Eq. (\ref{eq:QSDops})
\begin{eqnarray}
  \label{eq:QSDstate}
	&&\left|d\psi\right\rangle =  -\frac{i}{\hbar}H\left|\psi\right\rangle dt \nonumber\\
	&& + \sum_m\left(\left\langle L_m^\dagger\right\rangle_\psi L_m - \frac{1}{2}L_m^\dagger L_m - \frac{1}{2}\left\langle L_m^\dagger\right\rangle_\psi \left\langle L_m\right\rangle_\psi \right)	\left|\psi\right\rangle	dt \nonumber\\
	&& + \sum_m\left(L_m - \left\langle L_m \right\rangle \right) \left|\psi\right\rangle d\xi_m							
\end{eqnarray}
\noindent where $d\xi_m$ are independent complex differential random variables (Weiner increments), which satisfy the following statistical averages:
\begin{subequations}
  \label{eq:QSDstats}
	\begin{eqnarray}	
		E\left(d\xi_m\right) & = & 0\\
		E\left(d\xi_n d\xi_m\right) & = & 0\\
		E\left(d\xi_n^* d\xi_m\right) & = & \delta_{nm}dt						
	\end{eqnarray}
\end{subequations}

	\begin{figure*}
		\centering
		\includegraphics[width=0.70\textwidth]{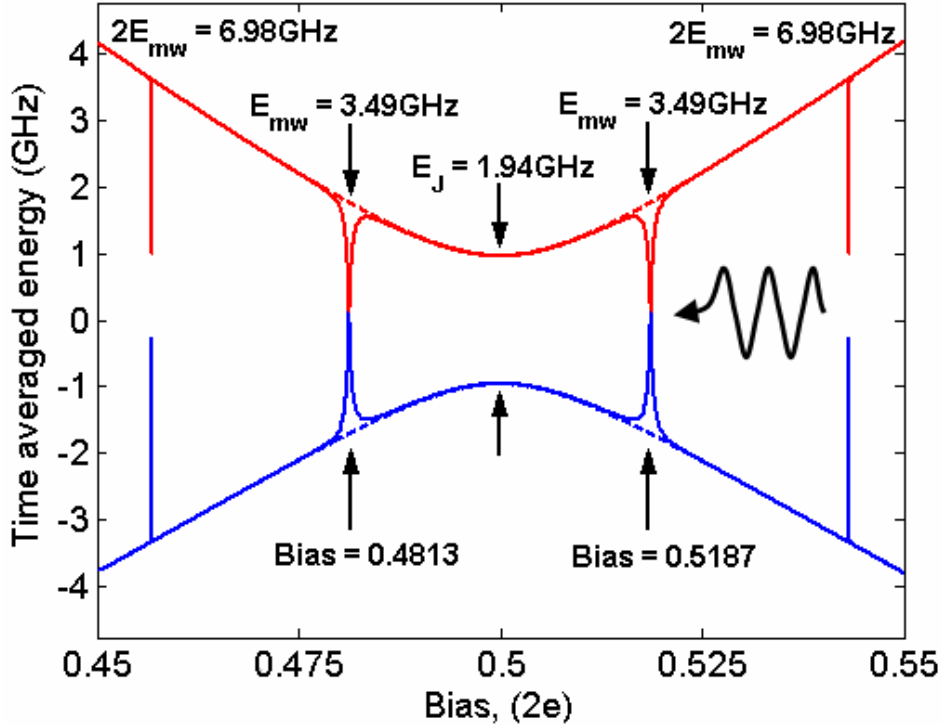}
		\caption{\label{fig:EnergyLevel} (Color online) Time averaged Floquet qubit energy levels, with one and two photon transitions (3.49GHz and 6.98GHz), following results focus on the region surrounding the single photon transition (3.49GHz) at Bias = 0.5187.  The solid upper and lower lines are the time averaged energies of the two states during microwave drive, the dashed upper and lower lines are the undisturbed energies.  The energy diagram clearly indicates the bias values near which the microwave transitions exist.  Time averaged Floquet energies are necessary to conveniently show the transitions as the microwave drive is a time varying field.}
		\label{fig:EnergyLevel}
	\end{figure*}

\noindent For the arbitrary index $m = 1,2$ we define two environment operators $L_m = L_1, L_2$, \cite{spiller:qsd}.
\begin{subequations}
	\label{eq:QSDops}
	\begin{eqnarray}
	  \label{eq:QSDopsA}	
		L_1 & = & \sqrt{\left(\bar{n}+1\right)\frac{\omega}{Q_D}}~a
		\\
		\label{eq:QSDopsB}
		L_2 & = & \sqrt{\bar{n}\frac{\omega}{Q_D}}~a^\dagger
		\\
		\label{eq:QSDopsC}
		\bar{n} & = & \frac{1}{\exp \left(\frac{\hbar\omega}{kT}\right)-1}	
	\end{eqnarray}
\end{subequations}

\noindent where $Q_D$ is the quality factor of the oscillator and is proportional to the dissipative elements, $\bar{n}$ is the average oscillator occupancy and $\omega$ is the angular frequency of the oscillator.  The oscillator occupancy is dependent on the thermal temperature of the environment which is assumed constant at T=10mK.


\subsection{\label{sec:sec3pB} Qubit quantum jumps}

To model the spontaneous photon emissions from the qubit as it undergoes a modified Schr\"{o}dinger evolution we adapt a quantum jump trajectory unravelling found in quantum optics \cite{wiseman:jumps}.  We allow the qubit to emit a photon into a dissipative reservoir (experimental cavity), in which it is immediately absorbed (An ideal measurement, to simulate a quantum jump).  If we apply continuous quantum measurements to the qubit we require two measurement operators (\ref{eq:QJops}), the subscript indicates the emission and absorption of a photon (i.e. a jump has occurred).
\begin{subequations}
	\label{eq:QJops}
  \begin{eqnarray}
    \label{eq:QJopsA}
		\Omega_0\left(dt\right) & = & 1 - \frac{i}{\hbar}Hdt - \frac{\gamma}{2}\sigma^\dagger\sigma dt
		\\ 
		\label{eq:QJopsB}
		\Omega_1\left(dt\right) & = & \sqrt{\gamma dt}~\sigma
  \end{eqnarray}
\end{subequations}

\noindent where $\sigma = \left| 0 \right\rangle \left\langle 1 \right|$, while $\gamma$ sets the jump rate so that $\gamma\left\langle \sigma^\dagger\sigma \right\rangle$ is the actual photon emission rate, ($\gamma = 0.05f_{osc}$ which implies 0.05 jumps per bias oscillator period, note that $\gamma$ differs for the two bias oscillator frequencies, see Table \ref{tab:components}).  Eq. (\ref{eq:QJopsA}) represents the continuous evolution of the qubit state over the interval $dt$, whereas Eq. (\ref{eq:QJopsB}) describes the collapse of the state when a quantum jump has occurred.
Over each simulation interval, $dt$, the state is updated via one of two Eqs. (\ref{eq:QJstate})
dependent on the random result of an instantaneous weak measurement, in normal evolution Eq. (\ref{eq:QJstateA}) is used, however if a jump occurs a `1' will be measured (photon emitted) and Eq. (\ref{eq:QJstateB}) should then be used and the state renormalised.
\begin{subequations}
	\label{eq:QJstate}
  \begin{eqnarray}
    \label{eq:QJstateA}
		\left|\psi_0\left(t + dt\right)\right\rangle & = & \Omega_0\left(dt\right) \left|\psi\left(t\right)\right\rangle  
		\\
		\label{eq:QJstateB}
		\left|\psi_1\left(t + dt\right)\right\rangle & = & \Omega_1\left(dt\right) \left|\psi\left(t\right)\right\rangle		
 \end{eqnarray}
\end{subequations}



	\begin{figure*}
		\centering
		\includegraphics[width=0.905\textwidth]{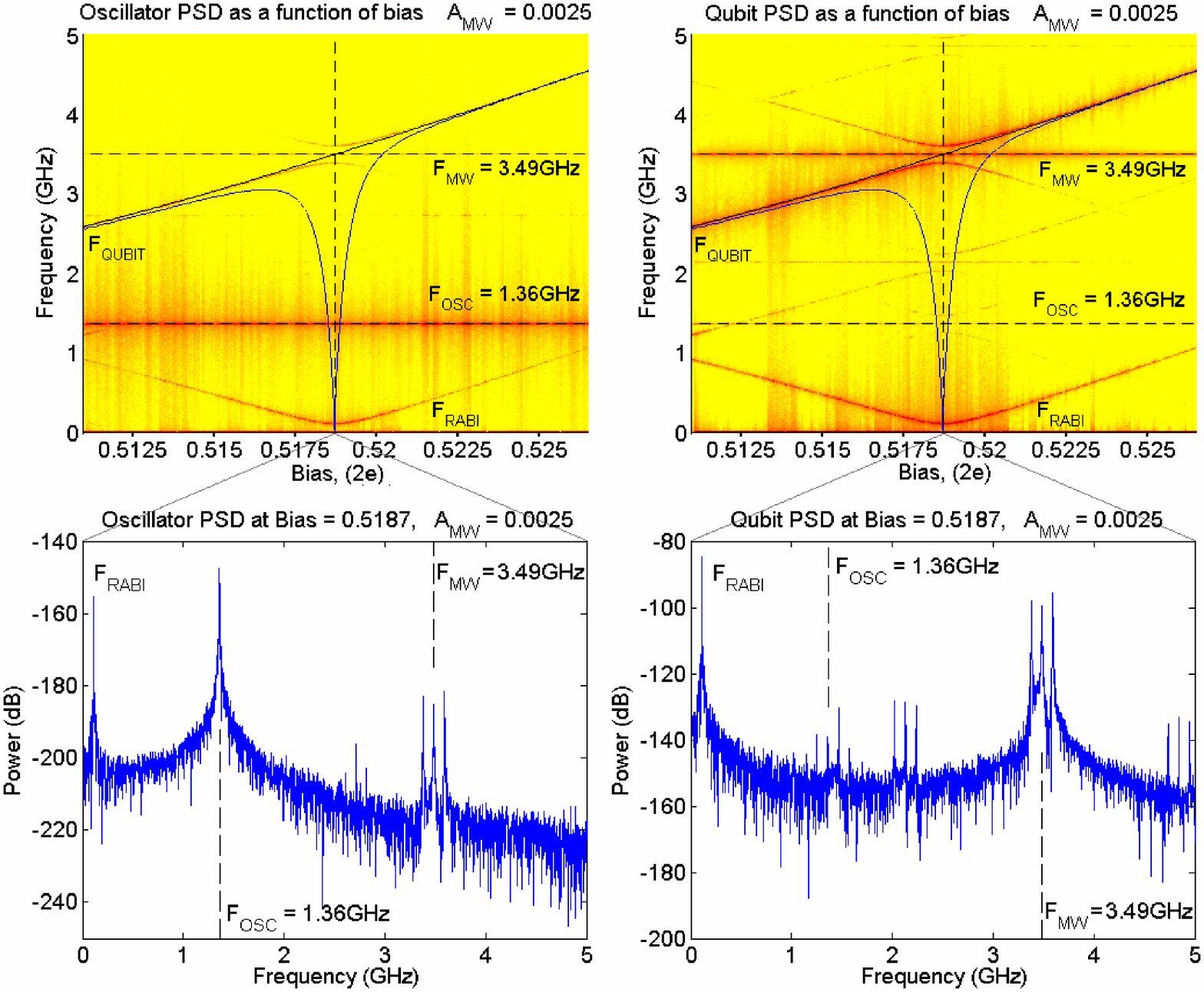}
		\caption{\label{fig:qubosc1d2d} (Color online) Oscillator and Qubit power spectra slices for Bias = 0.5187, using the low frequency oscillator circuit $f_{osc} = 1.36$GHz.  The solid lines overlay the energy level separations found in Fig.~\ref{fig:EnergyLevel}. ($\kappa = 5\times10^{-5}$).  As one would expect, the bias oscillator peak at 1.36GHz is clearly observed in the oscillator PSD, but only weakly in the qubit PSD.  Likewise the qubit Rabi frequency is found to be stronger in the qubit PSD.  However it is important to note that the qubit dynamics such as the Rabi oscillations are indeed coupled to the bias oscillator circuit and so can be extracted.  In addition, it is recommended to compare the layout of the most prominent features with Fig.~\ref{fig:BiasRamp}.}
		\label{fig:qubosc1d2d}
	\end{figure*}	

\section{\label{sec:sec4} ENERGY LEVEL STRUCTURE}

It is useful to be able to quantify several important qubit parameters, as the qubit behaviour is particularly sensitive to parameter variation.  These arise from the manufacturing techniques used, which have often been driven by industrial goals of delivering cheap yet inaccurate components.  It is important to understand that small variations in the capacitive coupling can have a significant effect on the system behaviour.  These errors could occur through a variety of real situations.  For example, overlaps between layers in the epitaxially grown structure, intrinsic parasitic capacitances such as fringe effects, or defects in the Josephson junctions \cite{Kerkhoff:defects}.   Unfortunately whilst these variations can be kept to a minimum by improved design and lithography, we cannot expect perfect structures, and once integrated and fabricated we cannot measure  every capacitance directly.  Instead, we are likely to require methods to remotely characterise the qubit in circuit. 

The qubit behaviour is partially characterised by the energy level diagram (Fig.~\ref{fig:EnergyLevel}) \cite{friedman:energygraph,scully:book}, which shows the time averaged (Floquet) energy levels for the microwave driven qubit \cite{clark:annals, Hochstadt:floquet}.  The Floquet energies are a useful illustration of where transitions can occur for a particular microwave drive frequency, and show how the qubit energy eigenvalues change as a function of the microwave and bias control fields.  However, this is not necessarily known and cannot be assumed to be identical for all qubits.

Indeed, techniques have been proposed to deduce the energy level structure from measurements of the increased peak noise power in the biasing circuit, caused by Rabi oscillations when the qubit is correctly biased (on resonance) with the injected microwave field \cite{griffith:charge,ralph:charge,ralph:flux}.  

	\begin{figure*}
		\centering
		\includegraphics[width=0.93\textwidth]{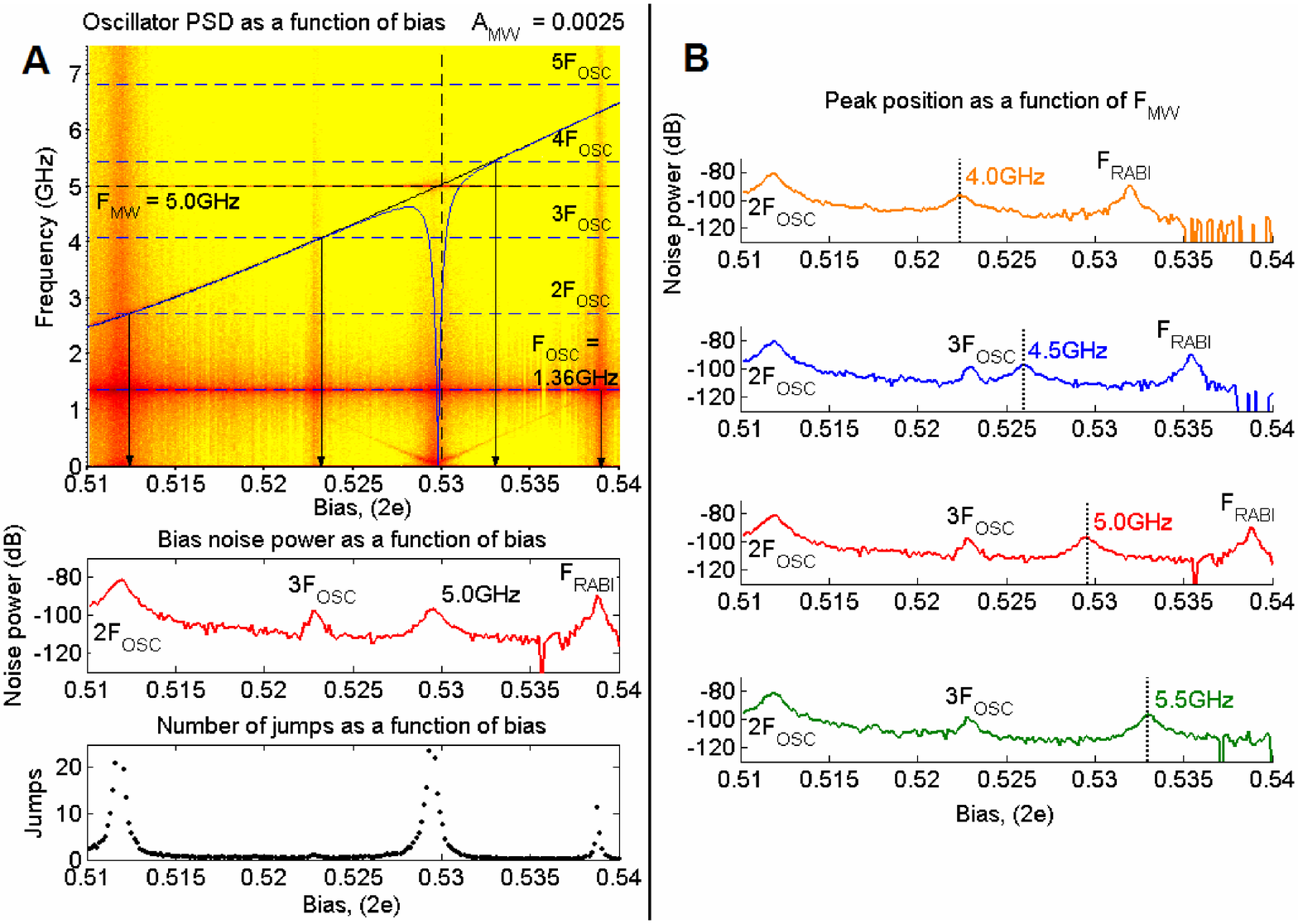}		
		\caption{\label{fig:Jumps} (Color online) (A) Oscillator power spectra when the coupled qubit is driven at $f_{mw} = 5.00$GHz. An increase in bias noise power ($f_{osc} = 1.36$GHz) can be observed when Rabi oscillations occur, the more frequent quantum jump noise couples back to the oscillator. (B) Bias noise power peak position changes as a function of $f_{mw}$, the microwave drive frequency.  Therefore, it is possible to probe the qubit energy level structure by using the power increase in the oscillator which is already in place, eliminating the need for additional measurement devices.  However, it should be noted that the surrounding oscillator harmonics may mask the microwave driven peak. ($\kappa = 1\times10^{-3}$).}
		\label{fig:Jumps}
	\end{figure*}

The system described in this paper is driven with a 3.49GHz microwave, which gives rise to a transition between the energy levels, the width of which becomes wider as the microwave amplitude ($A_{mw}$) is increased.
The single photon (3.49GHz) transition occurs symmetrically near two bias points (0.4812 and 0.5187).  We have chosen to operate about the upper bias point (0.5187), although identical features would be present near the lower point.
Two photon transitions occur at 6.98GHz, although the bias required to utilise these transitions are far above our parameter range, so their effects will be ignored.

The graph also clearly shows the expected avoided crossing caused by the Josephson junction, with an energy separation, $E_J$ (1.94GHz),
transitions cannot occur for microwaves driving below this frequency.  This is important, as only the 3.06GHz high frequency biasing oscillator should be sufficient to create the excitations that interfere with the microwave and cause the many strong and interesting features, unlike the 1.36GHz low frequency oscillator circuit.



\section{\label{sec:sec5} FREQUENCY SPECTRA}

In this section we consider the frequency spectrum of the system, or the power spectral density (PSD) which is an experimentally measurable quantity.  
The power spectral density  reveals the strength of the frequencies present in a time varying signal.  In this case the signals are voltages, which are calculated from the expectation values of qubit and oscillator charge.  The expectation values are obtained using the reduced density operator for the qubit and oscillator basis, applying the relevant charge operator and then tracing the result.\\
E.g. $\left\langle q \right\rangle = Tr\left\{ \hat{q}\rho_q \right\}$ where $\rho_q$ is the reduced density matrix for the qubit state.
\begin{subequations}
	\label{eq:Powers}
  \begin{eqnarray}
    \label{eq:QPower}
		P(Q) & = & 20\log_{10}{|FFT(V_Q)|} 
		\\
		\label{eq:qPower}
		P(q) & = & 20\log_{10}{|FFT(V_q)|} 
 \end{eqnarray}
\end{subequations}

\noindent where $V_Q = \left\langle Q \right\rangle /C_Q$ and $V_q = \left\langle q \right\rangle /C_q$.  $V_Q$ should be accessible though measurements on the biasing circuit, and $V_q$ (if required) can be measured using a Single Electron Transistor (SET).  FFT denotes the Fast Fourier Transform which is a computationally efficient implementation of the standard Fourier transform, to obtain the frequency spectra of the stochastic time evolutions $V_Q$ and $V_q$.

In the example figure (Fig.~\ref{fig:qubosc1d2d}), the control bias is varied from left to right for a low frequency oscillator circuit (1.36GHz).  For each bias point the simulation is reinitialised, the stochastic time evolution of the system density matrix is simulated over 1500 oscillator cycles.  Then the oscillator and qubit charge expectation values are extracted to obtain the power spectrum for each component, with a frequency resolution of 4.01MHz.  The power spectra for each time series are collated as an image such that the power axis is now represented as a colour, and the individual power spectra are vertical `slices' through the image.  The dominant frequency peaks become line traces,  therefore illustrating the various avoided crossings, mergeings and intersections.
The example figure shows the PSD `slice' at Bias $= 0.5187$, the broadband noise is readily apparent and is due to the discontinuous quantum jumps in the qubit.  The bias oscillator peak (1.36GHz) is most prominent in the oscillator PSD, as would be expected, but it is also present in the qubit PSD.  It should also be noted that most features are present in both the qubit and oscillator, including the noise which is generated by the quantum jumps and the quantum state diffusion processes.  Interestingly, the qubit PSD is significantly stronger than the oscillator PSD, however, a larger voltage is generated by the smaller charge due to the extremely small island capacitance, $(V_q = q/C_q)$.

In addition, the energy level separation (Fig.~\ref{fig:EnergyLevel}) has been overlaid as a solid curved line to illustrate the effect of the microwave drive.



In a previous paper \cite{ralph:flux}, a method was proposed by which the energy level structure of a charge qubit can be obtained from measurements of the peak noise in the bias/control oscillator, without the need of extra readout devices.  This was based on a technique originally proposed for superconducting flux qubits \cite{ralph:flux} but there are many similarities between the two technologies.  The oscillator noise peak is the result of broadband noise caused by quantum jumps in the qubit being coupled back to the oscillator circuit.  This increase in the jump rate becomes a maximum when the Rabi oscillations are at peak amplitude, this should only occur when the qubit is correctly biased and the microwave drive is driving at the transition frequency.  Therefore by monitoring this peak as a function of bias, we can associate a bias position with a microwave frequency equal to that of the energy gap, hence constructing the energy diagram (Fig.~\ref{fig:EnergyLevel}). 

The peak is approximately +10dB to +15dB above the background noise (Fig.~\ref{fig:Jumps}), which is set by the thermal fluctuations. Although it would be difficult to measure directly in an experiment, the electronic noise temperature at the first (low temperature) amplification stage should not be significantly above this thermal level or it would adversely affect the ability to measure quantum mechanical behaviour in the qubit.  Care should be taken to avoid detecting a fixed peak, these extra peaks are due to oscillator harmonics driving the qubit, however the far right peak is due to resonance between the Rabi oscillations and the bias oscillator, (Fig.~\ref{fig:Jumps}B).

The next two sections illustrate how the two control fields affect the features, by first using the above mentioned method for determining the energy level structure, it may be possible to determine the \textit{true} bias and microwave fields applied to the qubit: 
\begin{itemize}
	\item Section A investigates the effect of \textsl{sweeping the bias field} about the energy level transition, for a constant microwave frequency of 3.49GHz.
	\item Section B investigates the effect of \textsl{sweeping the microwave frequency}, at a constant bias of 0.5187.
\end{itemize}

In both sections, the example figures provided use the high frequency (3.06GHz) bias oscillator.  This higher frequency bias field requires a quantised oscillator model because the classical model is insufficient \cite{spiller:classosc}.  To illustrate the additional effects created when two frequencies are driving in close proximity, it is necessary to compare the following plots with Fig.~\ref{fig:qubosc1d2d}.  In addition the coupling, $\kappa$ is kept constant at $5\times10^{-5}$.



	\begin{figure*}
		\centering
		\includegraphics[width=1.0\textwidth]{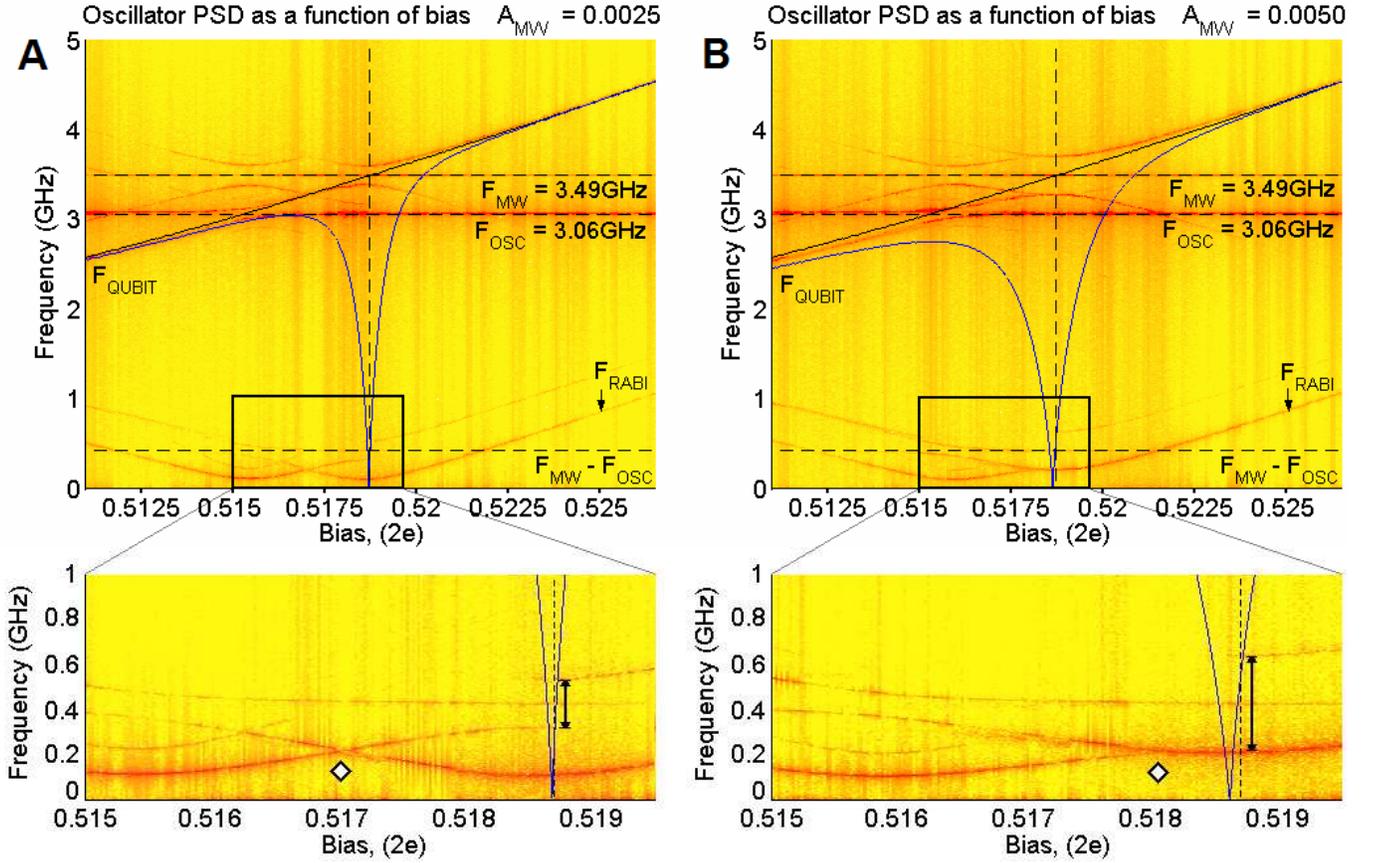}	
		\caption{\label{fig:BiasRamp} (Color online) Oscillator PSD as a function of bias, for microwave amplitudes $A_{mw}$ = 0.0025 (A) and $A_{mw}$ = 0.0050 (B).  The red lines track the positions (in frequency) of significant power spectrum peaks (+10dB to +15dB above background), the overlaid black and blue lines are the qubit energy and microwave transition (Fig.~\ref{fig:EnergyLevel}).  Unlike Fig.~\ref{fig:qubosc1d2d}, in these figures the 3.06GHz oscillator circuit can now drive the qubit (Fig.~\ref{fig:EnergyLevel}) and so creates excitations which mix with the microwave driven excitations creating a secondary splitting centred on $f_{mw}-f_{osc}$ (430MHz).  This feature contains the Rabi frequency information in the sidebands of the splitting, but now in a different and controllable frequency regime.  In addition, the intersection of the two differently driven excitations (illustrated in the magnified sections), opens the possibility of calibrating the biased qubit against a fixed engineered oscillator circuit, using a single point feature. ($\kappa = 5\times10^{-5}$).}
		\label{fig:BiasRamp}
	\end{figure*}	

\subsection{\label{sec:sec4pA} Bias field sweep}

Fig.~\ref{fig:BiasRamp} shows plots of the PSD for two different microwave amplitudes $A_{mw} = 0.0025$ (A) and $A_{mw} = 0.0050$ (B) driving at 3.49GHz with a high frequency bias oscillator of 3.06GHz.  

Although not shown, when the microwave field is removed there are still some effects due to the 3.06GHz high frequency oscillator causing excitations and a splitting in the qubit.  This is possible as the oscillator exceeds the Josephson junction frequency (1.94GHz) and so drives the qubit as per an injected microwave field.  Alternatively, if a low frequency oscillator (1.36GHz) is used, there are no excitations and the qubit transition frequency remains undisturbed, unless the qubit is biased on a harmonic of the oscillator, which would therefore exceed the Josephson frequency and start driving the qubit.

When the microwave is applied we notice interesting effects, arising from the additional interference of the `side bands' created by the 3.06GHz oscillator drive signal, which is now in close proximity to the 3.49GHz microwave field.  This relatively small detuning causes a multitude of splittings, similar to that of the Autler-Townes effect \cite{greentree:autler}.  However, this is a benefit as it creates many more features in the frequency spectrum, which in turn increases the frequency bands in which useful information can be obtained.  Although the features obtained are clearly related, the presence of multiple frequencies (mixtures, splittings and harmonics) in different regions of frequency provides different experimental `windows' within which to explore the behaviour of Josephson charge devices.

The mixing of the Rabi oscillations with the microwave field causes `side bands' surrounding the drive signal.  It is important to note that the high frequency behaviour surrounding the drive signals is replicated at $f_{mw}-f_{osc}=$ 430MHz, and this area of interest has been enlarged, indeed, this should be an easily accessible frequency range.  This frequency regime can be adjusted on to an experimental bandwidth by changing the oscillator frequency, and hence the aforementioned separation, $f_{mw}-f_{osc}$.

In addition, within the PSD there exists a point of interest ($\diamondsuit$) where the microwave and oscillator excitations intersect.  As the hyperbolic shape of the oscillator excitations remains quite constant for low amplitude microwaves, the intersection point will mainly be governed by the amplitude and frequency of the microwave.  It can be seen in plot A, when the microwave is driving at a similar amplitude to the oscillator signal ($A_{mw} = 0.0025$), the intersection occurs halfway between the two signals.  

This is a useful feature that may be of direct benefit to experimental investigators.  The application of microwaves to this sort of solid state system is normally via a frequency dependent coupling.  If multiple frequencies are to be used in an experiment, each will need to be characterised.  This feature will provide an independent measure for the amplitude of the microwaves which actually couple to the device at different frequencies.  It should allow the amplitude of the microwaves reaching the device to be kept constant even though the coupling changes with frequency.

In addition, it may be possible to characterise the oscillator in terms of microwave properties, and furthermore estimate the effects or significance of the oscillator energy transition.  Indeed, for high microwave amplitudes $(A_{mw} > 0.0050)$, the microwave transition width eventually increases to enclose the oscillator transition, then the features merge.



	\begin{figure*}
		\centering	
		\includegraphics[width=1.0\textwidth]{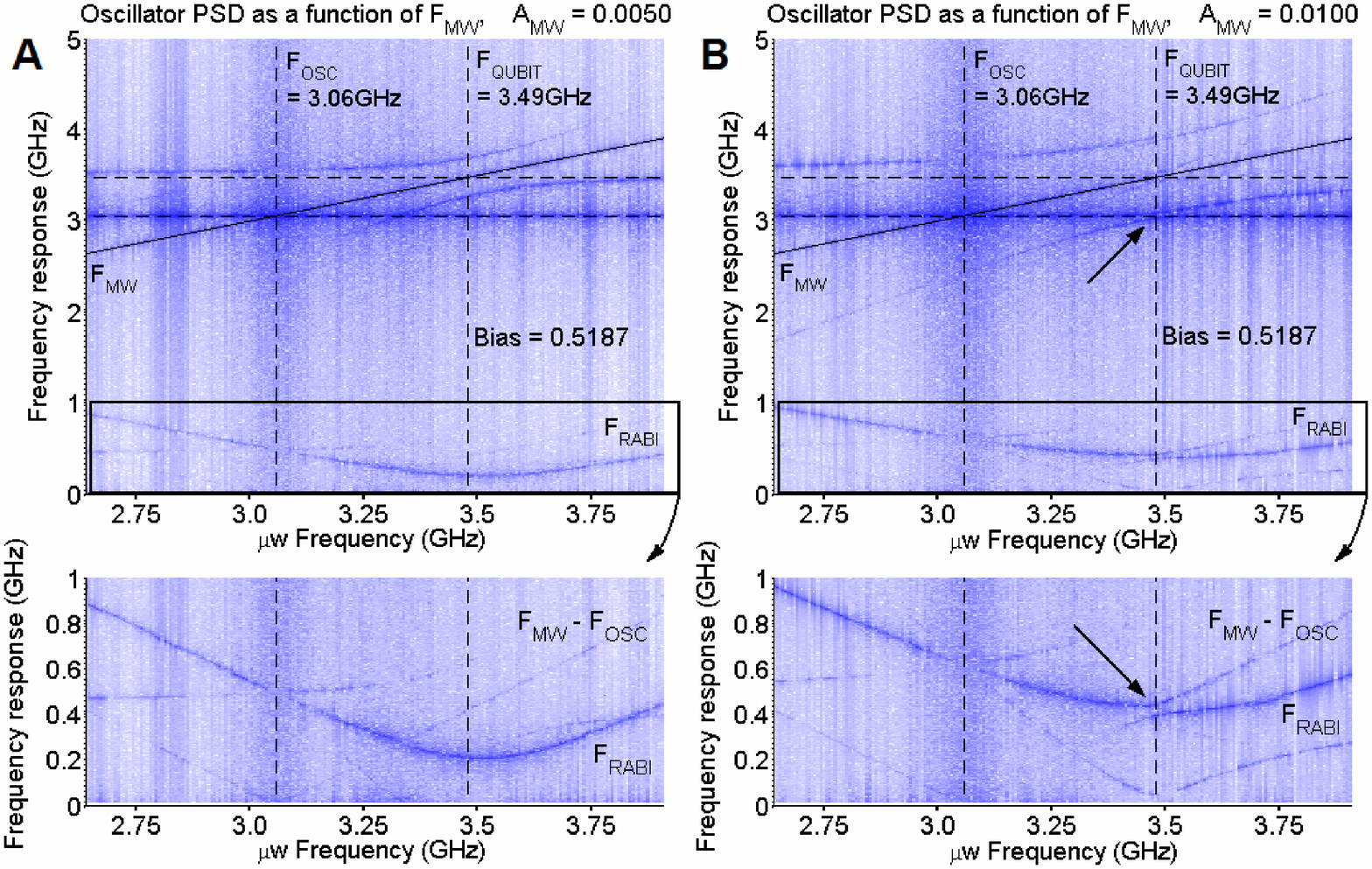}	
		\caption{\label{fig:mwRamp} (Color online) Oscillator PSD as a function of the applied microwave drive frequency $f_{mw}$, for microwave amplitudes $A_{mw}$ = 0.0050 (A) and $A_{mw}$ = 0.0100 (B).  It is important to notice that there are now two frequency axes per plot, a drive (H) and a response (V).  Of particular interest is the magnified section which shows clearly the distinct secondary splitting in the sub-GHz regime.  This occurs due to a high frequency interaction seen in the upper plots, where the lower Rabi sideband of the microwave drive passes through the high frequency oscillator signal.  The maximum splitting occurs when the Rabi amplitude is a maximum, hence this is observed for a very particular combination of bias and drive, which is beneficial for charactering the qubit.  Most importantly, this would not be observed with a conventional low frequency oscillator configuration as the $f_{mw}-f_{osc}$ separation would be too large for the Rabi frequency. ($\kappa = 5\times10^{-5}$).}
	  \label{fig:mwRamp}
	\end{figure*}	

\subsection{\label{sec:sec4pB} Microwave frequency sweep}

Fig.~\ref{fig:mwRamp} is presented in a similar manner as Fig.~\ref{fig:BiasRamp}.  However there are now two frequency axes: the horizontal axis represents the frequency of the applied microwave drive field, and the vertical axis is the frequency response.  It should be remembered that the microwave frequency axis is focused near the qubit transition frequency ($f_{qubit}\approx$ 3.49GHz) and the diagonally increasing line is now the \textit{microwave frequency}.

The most interesting features in question are secondary splittings, which occur only when the lower Rabi side-band passes through the oscillator peak, splitting it.  The oscillator frequency split is somewhat obscured by the large oscillator peak, however the low frequency splitting centred on $f_{mw}-f_{osc}=$ 430MHz is clearly visible in the magnified section, as indicated by the arrows.
 
We find the particular combination required to cause maximum splitting only occurs when $f_{Rabi} = f_{mw}-f_{osc}$ \textbf{and} the Rabi oscillations have maximum amplitude, (i.e. when the microwave is in resonance with the bias setting).  Hence the microwave amplitude and a feature in frequency space can be related.  The traces created by these splittings can be observed at a given applied microwave frequency, the traces expand and contract dependant on the proximity to the splitting.  Tracking the expansion of these traces it may be possible to tune the microwave amplitude for the desired Rabi frequency.

This could be applied to calibrate the microwave waveguide, in which one could determine the effective microwave applied to the qubit in terms of the microwave amplitude applied at the external end of the waveguide.  This is important as the characteristics of the waveguide, such as impedance, are frequency dependant.  If the separation of the two known frequencies is fixed, (the oscillator and the microwave), we know by maximising the secondary splitting using the microwave amplitude that the Rabi side-band must now be equal to the well defined oscillator-microwave frequency separation, and therefore the external microwave amplitude that caused it.  In addition, the maximum splitting solution should indicate a correctly biased system.
If the oscillator were of a much lower frequency (1.36GHz) then only the Rabi oscillations and Rabi side-bands would appear (Fig.~\ref{fig:qubosc1d2d}), and none of these additional interactions would exist.  The extra features caused by this small detuning is an important point to reinforce as it links the microwave, oscillator, Rabi oscillations and bias field together.
\section{\label{sec:sec6} Conclusion}

In this paper we have presented features that exist within the power spectra of a charge qubit. The qubit is coupled via a single Josephson junction to a (fully quantised) oscillator circuit, which provides a bias voltage and a means of measurement.  A quantum trajectory approach has been used, with the quantum state diffusion unravelling of the master equation simulating a dissipative oscillator, and a quantum jumps unravelling simulates spontaneous emissions from the qubit to a lossy reservoir.

A backreaction technique is tested with a quantised oscillator to allow the energy level separation of a charge qubit to be measured in-situ and with the minimum of measuring circuitry and without a separate qubit readout device.  However care should be taken to avoid detecting the higher oscillator harmonics.   We also report additional features that are created in the power spectrum of the qubit (and the coupled oscillator) when the oscillator resonant frequency and microwave frequency are similar (3.06GHz and 3.49GHz respectively).  These features are predominantly discontinuous frequency splittings akin to the Autler-Townes effect in quantum optics, and the intersections of frequency peaks as functions of the control fields.  These effects are also present at low (sub GHz) frequencies which are expected to be more easily accessible by experiment than the original 2 to 4 GHz frequency range.  This frequency range can be  moved into an experimental bandwidth by adjusting the oscillator frequency.  

The close proximity of a second drive field (the oscillator), to the microwave drive adds information to the noise spectra and allows access to other features, by searching for frequency separations and discontinuities we can identify bias settings and new amplitudes from artefacts in the frequency domain. 
\\

\begin{center}
\textbf{Acknowledgements}
\end{center}

The authors would like to thank Drs. P. J. Meeson, M. J. Everitt and J. H. Cole for helpful comments and suggestions.\\
\indent E. J. Griffith is supported by a Department of Electrical Engineering and University of Liverpool scholarship.
A. D. Greentree is supported by the Australian Research Council, the Australian government and by the US National Security Agency (NSA), Advanced Research and Development Activity (ARDA) and the Army Research Office (ARO) under contract number W911NF-04-1-0290.

\bibliography{AutlerTownesbib_proof}


\end{document}